\documentclass[amsmath,aps,showpacs,a4paper,10pt]{revtex4}

 \usepackage{epsf}
 \usepackage{graphicx}    

 \usepackage{rotating}    

\begin{document}

 \newcommand{\be}[1]{\begin{equation}\label{#1}}
 \newcommand{\ee}{\end{equation}}
 \newcommand{\bea}{\begin{eqnarray}}
 \newcommand{\eea}{\end{eqnarray}}
 \def\disp{\displaystyle}

 \def\gsim{ \lower .75ex \hbox{$\sim$} \llap{\raise .27ex \hbox{$>$}} }
 \def\lsim{ \lower .75ex \hbox{$\sim$} \llap{\raise .27ex \hbox{$<$}} }

 \begin{titlepage}

 \begin{flushright}
 arXiv:1912.03466
 \end{flushright}

 \title{\Large \bf The Possible Electromagnetic Counterparts
 of\\ the First High-Probability NSBH Merger\\ LIGO/Virgo S190814bv
 \vspace{2mm}}

 \author{Hao~Wei\,}
 \email[\,Corresponding author; email address:\ ]{haowei@bit.edu.cn}
 \affiliation{\vspace{1mm} School of Physics,
 Beijing Institute of Technology, Beijing 100081, China}

 \author{Minzi~Feng\,}
 \affiliation{\vspace{1mm} Key Laboratory of Particle Astrophysics,
 Institute of High Energy Physics, Chinese Academy of Sciences,
 Beijing 100049, China
 \\ University of Chinese Academy of Sciences, Beijing 100049, China}

 \begin{abstract}\vspace{6mm}
 \centerline{\bf ABSTRACT}\vspace{2mm}
 LIGO/Virgo S190814bv is the first high-probability neutron star --
 black hole (NSBH) merger candidate, whose gravitational waves (GWs)
 triggered LIGO/Virgo detectors at 21:10:39.012957 UT, 14 August 2019.
 It has a probability $>99\%$ of being an NSBH merger, with a low false
 alarm rate (FAR) of 1 per 1.559e+25 years. For an NSBH merger,
 electromagnetic counterparts (especially short gamma-ray bursts (GRBs))
 are generally expected. However, no electromagnetic counterpart has
 been found in the extensive follow-up observing campaign. In the present
 work, we propose a novel explanation to this null result. In our scenario,
 LIGO/Virgo S190814bv is just a GW mirror image of the real NSBH merger
 which should be detected before 14 September 2015, but at that time we
 had no ability to detect its GW signals. The electromagnetic counterparts
 associated with the real NSBH merger should be found in the archive data
 before 14 September 2015. In this work, we indeed find 9 short GRBs as
 the possible electromagnetic counterparts.
 \end{abstract}


 \maketitle

 \end{titlepage}

 \renewcommand{\baselinestretch}{1.0}


\section{Introduction}\label{sec1}

The first direct detection of the gravitational wave (GW) event on 14
 September 2015~\cite{Abbott:2016blz,TheLIGOScientific:2016src} opened a
 new window in physics and astronomy. Since the first GW event,
 GW150914, one can study the universe by using GWs, in addition
 to the traditional means mainly based on electromagnetic radiations.
 In the case of binary black hole (BBH) merger (like GW150914),
 only GWs are expected. However, if at least one of the binary
 compact objects is a neutron star, electromagnetic radiations will be
 also expected in addition to GWs. In fact, the first detection of binary
 neutron star (BNS) merger GW170817~\cite{TheLIGOScientific:2017qsa,
 GBM:2017lvd,Monitor:2017mdv,GW170817det} began a new era of
 multi-messenger astronomy. Through GWs, LIGO/Virgo detected the BNS merger
 GW170817 at 12:41:04 UTC, 17 August 2017~\cite{TheLIGOScientific:2017qsa,
 GW170817det}. Only $\sim 1.7\,{\rm s}$ later, the Fermi Gamma-ray Burst
 Monitor (GBM) independently detected a short gamma-ray burst
 GRB170817A~\cite{GBM:2017lvd,Monitor:2017mdv,GRB170817A}. Then, an
 extensive observing campaign~\cite{G298048} was launched across the
 electromagnetic spectrum leading to the discovery of a bright optical
 transient (SSS17a, or, AT 2017gfo) in NGC 4993~\cite{GBM:2017lvd}. Such
 an extensive observing campaign~\cite{G298048} covering GW, gamma-ray,
 X-ray, radio, ultraviolet, optical, near-infrared, infrared,
 neutrino observations and so on is very impressive in scientific history.

It is natural to expect another one. During the first observing run (O1,
 from 2015-09-12 to 2016-01-19) of LIGO/Virgo, GWs from three BBH mergers
 were detected; and during the second observing run (O2, from 2016-11-30
 to 2017-08-25), GWs from seven BBH mergers and one BNS merger (GW170817)
 were detected~\cite{LIGOScientific:2018mvr,O1O2}. Unfortunately, NO further
 BNS mergers beside GW170817 were detected, and also NO neutron star --
 black hole (NSBH) mergers were detected~\cite{LIGOScientific:2018mvr,O1O2}.

LIGO/Virgo began the third observing run (O3) on 1 April 2019, which is
 scheduled to end on 30 April 2020, and the KAGRA detector may join the
 later part of the O3 run as commissioning progress permits~\cite{O3andon,
 O3plan}. The official candidate event database is GraceDB~\cite{GraceDB}.
 36 GW detection candidates were found as of 1 December
 2019~\cite{GraceDBO3}. Among them, LIGO/Virgo S190814bv is the first
 high-probability NSBH merger candidate~\cite{S190814bv,S190814bvNotice,
 S190814bvGCN}, whose GWs triggered LIGO/Virgo detectors
 at 21:10:39.012957 UT, 14 August 2019~\cite{S190814bvNotice}.
 It has a probability $>99\%$ of being an NSBH merger, with a low false
 alarm rate (FAR) of 1 per 1.559e+25 years, at a distance of
 $267\pm 52\,{\rm Mpc}$~\cite{S190814bv,S190814bvNotice,S190814bvGCN}.
 Unlike most GW events, S190814bv was well localized to a small area
 of 23 (5) ${\rm deg}^2$ at $90\%$ ($50\%$)
 confidence~\cite{S190814bv,S190814bvGCN}, mainly thanks to its favorable
 location in the sky with respect to the antenna pattern of the three
 LIGO/Virgo GW detectors~\cite{Andreoni:2019qgh}, while it was detected by
 all the three instruments H1, L1 and V1~\cite{S190814bv}. The
 classification ``\,NSBH\,'' means that the lighter component has a
 mass $< 3M_\odot$ and the heavier component has a mass
 $> 5M_\odot$~\cite{Andreoni:2019qgh,Gomez:2019tuj,Dobie:2019ctw}. According
 to the physical upper-limit for neutron star mass~\cite{Ozel:2016oaf} (see
 also e.g.~\cite{Cromartie:2019kug,Zhang:2019bhn}), it is commonly believed
 that the lighter one with a mass $< 3M_\odot$ is probably a neutron star,
 while the heavier one with a mass $> 5M_\odot$ is probably a black hole.

For a long time, NSBH mergers were theorized as potential sites
 of $r$-process nucleosynthesis~\cite{Lattimer:1974slx}, leading to
 GWs detectable by laser interferometers like
 LIGO/Virgo~\cite{Abadie:2010cf}, associated with the possible
 electromagnetic counterparts including short gamma-ray bursts (GRBs),
 optical and radio afterglows, as well as day-long optical transients
 (kilonovae)~\cite{Metzger:2011bv}. In particular, short GRBs are promising
 and have long been suspected on theoretical grounds to arise from NSBH or
 BNS mergers~\cite{Berger:2013jza}.

Therefore, another extensive follow-up observing
 campaign~\cite{S190814bvGCN} began immediately after the public alert
 of LIGO/Virgo S190814bv. Unfortunately, NO evidence for a coincident
 short GRB was found~\cite{S190814bvGCN}, and optical, near-infrared,
 radio observations found numerous candidate counterparts but they were
 quickly ruled out~\cite{S190814bvGCN,Andreoni:2019qgh,Gomez:2019tuj,
 Dobie:2019ctw}. As of 1 December 2019, NO electromagnetic counterpart
 of LIGO/Virgo S190814bv has been found~\cite{S190814bvGCN,S190814bv}.
 This extensive observing campaign ended with nothing.

How to understand this null result? At first, there exists the possibility
 that LIGO/Virgo S190814bv might be classified as ``\,MassGap\,'' instead of
 NSBH, with a low probability $<1\%$~\cite{S190814bv,S190814bvNotice}.
 A MassGap system refers to a binary where the lighter companion has a
 mass $3M_\odot < M < 5M_\odot\,$, and no material is expected to be
 ejected, so that the merger is unlikely to produce electromagnetic
 emission~\cite{Andreoni:2019qgh,Gomez:2019tuj}. On the other hand,
 the actual upper limit for neutron star mass is not well constrained
 in fact, and hence S190814bv might actually be a
 BBH merger~\cite{Andreoni:2019qgh,Gomez:2019tuj,Dobie:2019ctw}, while the
 lighter component could be a low-mass black hole (e.g. a primordial black
 hole). For a BBH merger, no electromagnetic counterpart is expected.
 Another possibility is that the tidal breakup of the neutron star near the
 black hole~\cite{Lattimer:1974slx,Berger:2013jza} has not happened, and
 the black hole could have swallowed the neutron star in one clean gulp
 with little left to see~\cite{SA}. However, all the above explanations
 have their own difficulties. The probability of the first one is low,
 namely $<1\%$~\cite{S190814bv,S190814bvNotice}. The second one will
 impact the theories of neutron star and black hole about the upper limit
 for neutron star mass and the lower limit for black hole mass (as well as
 the theory of primordial black hole if it is invoked). The third one will
 impact the theory of NSBH merger~\cite{Lattimer:1974slx,Berger:2013jza}.
 By now, the parameters of the binary LIGO/Virgo S190814bv (e.g. mass
 and spin) are not publicly available, and hence the status is still
 unclear.

While the BNS merger GW170817 opened a new era of multi-messenger astronomy,
 could the high-probability NSBH merger candidate S190814bv be a stage for
 crazy ideas? In the present work, we try to propose a novel explanation.
 In our scenario, LIGO/Virgo S190814bv corresponds to a real NSBH merger
 associated with electromagnetic counterparts. However, the GW event
 LIGO/Virgo S190814bv detected on 14 August 2019 is not the NSBH merger
 itself. The real NSBH merger associated with electromagnetic counterparts
 should be detected many years before 14 September 2015, but at that time
 we had no ability to detect its GW signals, while its electromagnetic
 counterparts might be recorded in the archive data. LIGO/Virgo S190814bv
 is just a mirror image of this real NSBH merger, and we saw this mirror
 image many years later through GWs. However, the mirror imaging mechanism
 only works for GWs, not for electromagnetic signals. Therefore, we cannot
 find the electromagnetic counterparts while we detected GWs of LIGO/Virgo
 S190814bv on 14 August 2019. In fact, the corresponding electromagnetic
 counterparts should be found in the archive data before 14 September 2015.

In Sec.~\ref{sec2}, we will discuss the mirror imaging mechanism for
 GWs in detail. In Sec.~\ref{sec3}, we will discuss the method
 to find the electromagnetic counterparts in the archive data before
 14 September 2015. And then, in the archive data we indeed find several
 short GRBs probably associated with LIGO/Virgo S190814bv. In
 Sec.~\ref{sec4}, a brief summary will be given.


\section{A mirror imaging mechanism for GWs}\label{sec2}

Recently, a novel mirror imaging mechanism for GWs was proposed
 in~\cite{Wei:2019sqw}. This mirror imaging mechanism is related to
 superconductivity. In~\cite{Minter:2009fx,Chiao:2009tn,
 Chiao:2007pe,Chiao:2017rfe}, superconducting film was predicted to be a
 highly reflective mirror for GWs (see also e.g.~\cite{Quach:2015qwa,
 Inan:2017qdt,Inan:2017ixx}). Following~\cite{Minter:2009fx,Chiao:2009tn,
 Chiao:2007pe,Chiao:2017rfe,Quach:2015qwa,Inan:2017qdt,Inan:2017ixx}, let
 us try to give a perceptive picture for this GW mirror reflection. As a
 ripple in spacetime, the effect of a passing GW is mainly to make the
 particles follow the distortion in spacetime and then float (freely
 fall). In the superconductor, according to the well-known
 Bardeen-Cooper-Schrieffer (BCS) theory of superconductivity, negatively
 charged Cooper pairs will be formed. The Cooper pairs in the BCS ground
 state are in an exactly zero-momentum eigenstate, and hence their positions
 are completely uncertain, namely their trajectories are completely
 delocalized, due to the Heisenberg uncertainty relation for momentum
 and position. This quantum delocalization of Cooper pairs is protected
 from the localizing effect of decoherence by the BCS energy gap. Thus,
 Cooper pairs cannot undergo free fall along with the positively charged
 ions and normal electrons. In the presence of a GW, Cooper pairs of a
 superconductor undergo non-geodesic motion relative to the geodesic
 motion of its ionic lattice. So, Cooper pairs cannot respond at all to
 the passage of a GW, in contrast to the positive ions. This non-geodesic
 motion leads to the existence of mass and charge supercurrents inside
 the superconducting film. The generation of supercurrents by a GW has
 an important consequence, namely the electrical polarization of the
 superconductor. The resulting separation of oppositely signed charges
 leads to a huge Coulomb force that strongly opposes the tidal force of
 the incoming GW. Thus, this incoming GW is expelled, and then reflected.
 Such an effect in a superconductor was called the ``\,Heisenberg-Coulomb
 effect\,''. We refer to~\cite{Minter:2009fx,Chiao:2009tn,
 Chiao:2007pe,Chiao:2017rfe,Quach:2015qwa,Inan:2017qdt,Inan:2017ixx}
 for technical details.

Due to historical reasons, GWs at microwave frequencies were
 mainly considered in~\cite{Minter:2009fx,Chiao:2009tn,Chiao:2007pe,
 Chiao:2017rfe,Quach:2015qwa,Inan:2017qdt,Inan:2017ixx}, far before the first
 direct GW detection by LIGO. However, the arguments in~\cite{Minter:2009fx,
 Chiao:2009tn,Chiao:2007pe,Chiao:2017rfe,Quach:2015qwa,Inan:2017qdt,
 Inan:2017ixx} are also applicable for GWs at much lower frequencies (e.g.
 $1\,$Hz to $100\,$Hz GWs detectable for LIGO/Virgo). In fact, it was
 argued that for the incident GWs at angular frequencies
 $\omega\ll 10^{16}\,{\rm rad/s}$ (equivalently $f\ll 10^{15}\,{\rm Hz}$),
 the reflectivity ${\cal R}_{\rm G}$ is extremely close to
 $100\%$~\cite{Minter:2009fx,Chiao:2009tn,Chiao:2007pe,Chiao:2017rfe}. Thus,
 the mirror reflection of GWs from superconducting film is possible.
 Unfortunately, this type of experiment on earth for laboratory-scale
 superconducting mirrors of GWs is in fact very difficult to achieve.

\begin{center}
 \begin{figure}[tb]
 \centering
 \vspace{-10mm}  
 \includegraphics[width=0.82\textwidth]{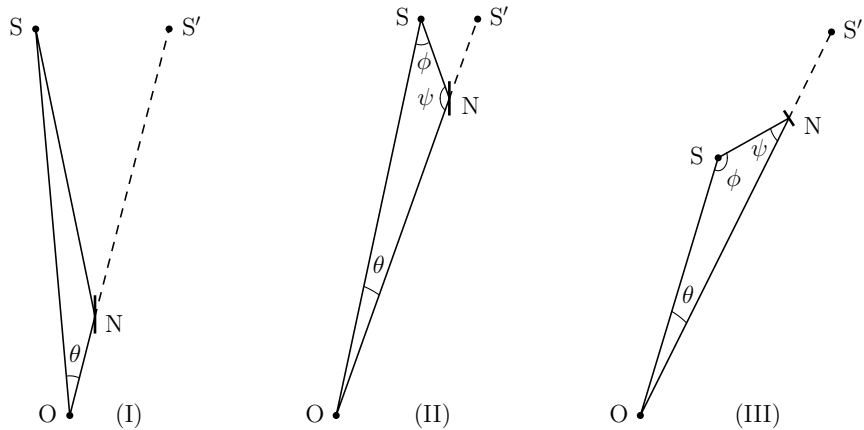}
 \caption{\label{fig1} Panel (I) corresponds to case (2a)
 of~\cite{Wei:2019sqw}, namely the neutron star N as the mirror for GWs
 is much closer to the observer O on earth, with luminosity distance
 $d_{\rm ON}\ll d_{\rm SN}$. Panels (II) and (III) correspond to cases
 (3a) and (3b) of~\cite{Wei:2019sqw}, namely the neutron star N
 as the mirror for GWs is much closer to the GW source S, with
 luminosity distance $d_{\rm SN}\ll d_{\rm ON}$. $\rm S^\prime$ is the
 mirror image of the GW source S. All plots are not to scale. See the
 text and~\cite{Wei:2019sqw} for details.}
 \end{figure}
 \end{center}

In~\cite{Wei:2019sqw}, it was proposed that the GW mirror reflection might
 happen in the sky. As is well known, neutron stars exhibit
 superconductivity and superfluidity (see e.g.~\cite{Lattimer:2004pg,
 Baym:1978jf,Weber:2004kj,Page:2010aw,Baldo:2002ju}). They could play
 the role of plane mirrors (superconducting films) for GWs in the universe.
 In~\cite{Wei:2019sqw}, it was predicted that there are two types of GW
 mirror imaging phenomena caused by the neutron star located in Milky Way
 or the same host galaxy of the GW source, which might be detected within a
 life period of man (namely the time delay $\Delta t$ can be a few years
 to a few tens of years). To be self-contained, we reproduce the plots
 (2a), (3a) and (3b) of~\cite{Wei:2019sqw} as the plots (I), (II) and (III)
 in Fig.~\ref{fig1} (not to scale). The observer O on earth firstly
 detects a GW signal from the GW source S directly, and after a time delay
 $\Delta t = (d_{\rm SN}+d_{\rm ON}-d_{\rm OS})/c$ the observer O will
 detect a secondary GW signal from the mirror image $\rm S^\prime$.
 The neutron star N plays the role of mirror for GWs. $c$ is the speed
 of light. Note that the luminosity distance of GW source $d_{\rm OS}$
 is usually huge ($\,{\cal O}(10^2)$ Mpc or larger). In case (I), the
 neutron star N can be located in the same host galaxy of O (namely
 Milky Way), with luminosity distance $d_{\rm ON}\ll d_{\rm SN}$, and
 hence $c\Delta t\simeq\left(1-\cos\theta\right)d_{\rm ON}$. In cases
 (II) and (III), the neutron star N can be located in the same host galaxy
 of S, with luminosity distance $d_{\rm SN}\ll d_{\rm ON}$, and hence
 $c\Delta t\simeq\left(1-\cos\phi\right)d_{\rm SN}$ and
 $c\Delta t\simeq\left(1+\cos\psi\right)d_{\rm SN}$, respectively. Note
 that the neutron star N as the mirror for GWs is not the neutron star in
 the binary NSBH or BNS. As is shown in~\cite{Wei:2019sqw}, only in these
 three cases can the time delay $\Delta t$ be within the lifespan of a human
 (namely a few years to a few tens of years). In the other cases considered
 in~\cite{Wei:2019sqw}, $\Delta t$ will be far beyond the lifespan of a
 human. Obviously, this novel imaging mechanism for GWs is different
 from the gravitational lens, or other phenomena such as the Poisson-Arago
 spot~\cite{Hongsheng:2018ibg}. We refer to~\cite{Wei:2019sqw} for details.


 \vspace{-10mm}  


\section{The possible electromagnetic counterparts}\label{sec3}

Let us consider the high-probability NSBH merger candidate LIGO/Virgo
 S190814bv in light of the mirror imaging mechanism for GWs mentioned
 above. It is possible that the corresponding NSBH merger ``\,S\,''
 associated with electromagnetic counterparts is real, but its GW and
 electromagnetic signals arrived at earth many years before 14
 September 2015. At that time we had no ability to detect its
 GW signals, while its electromagnetic counterparts might be recorded
 in the archive data. LIGO/Virgo S190814bv is nothing but the mirror
 image ``$\,\rm S^\prime\,$''. The structure of a neutron star
 is complicated~\cite{Lattimer:2004pg,Baym:1978jf}. When the GWs and
 electromagnetic radiations from the NSBH merger ``\,S\,'' arrive at the
 non-superconducting surface of the neutron star, most of the
 electromagnetic radiations are absorbed, while the surface of the neutron
 star is almost transparent to GWs. Then, the incident GWs will
 be reflected by the inner superconducting structure of the neutron star
 (which can be safely regarded as plane mirror (superconducting film)
 for GWs, due to the huge luminosity distance of the GW source and the small
 size of the neutron star). So, only the reflected GWs can be detected after
 a time delay $\Delta t$, but no electromagnetic counterparts can be
 found at the same time. This is the very case of LIGO/Virgo~S190814bv.

If the real NSBH merger corresponding to LIGO/Virgo S190814bv and its
 electromagnetic counterparts does not happen before 14 September 2015,
 according to the mirror imaging mechanism mentioned above, LIGO/Virgo
 should detect another GW signal with almost same intensity and wave
 form of LIGO/Virgo S190814bv between 14 September 2015 and 14 August
 2019. However, this did not happen in fact. So, we must assume that
 the real NSBH merger and its electromagnetic counterparts can only be
 found before 14 September 2015 (the day when LIGO directly detected
 the first GW signal).

As the next step, it is natural to find the possible electromagnetic
 counterparts associated with the real NSBH merger ``\,S\,'' in the
 archive data before 14 September 2015. The key is to find the neutron
 star ``\,N\,'' as the mirror for GWs at first. As mentioned above,
 there are two types of the location of the neutron star ``\,N\,'' as
 the mirror for GWs. So, we consider case (I) and cases (II), (III)
 in the following two subsections, respectively.

 \vspace{-4mm}  


\subsection{Case (I)}\label{sec3a}

Since almost all known neutron stars are located in Milky Way and the
 nearby Magellanic Clouds, we firstly consider case (I) of
 Fig.~\ref{fig1} (i.e. the case (2a) of~\cite{Wei:2019sqw}), namely
 the neutron star ``\,N\,'' is located in Milky Way (and Magellanic
 Clouds). From the left panel of Fig.~\ref{fig1}, it is easy to see that
 the neutron star ``\,N\,'' as the mirror for GWs is in the same direction
 of the image ``$\,\rm S^\prime\,$'' (namely LIGO/Virgo S190814bv) in this
 case. To our best knowledge, there are about 3000 known neutron stars
 by now~\cite{Manchester:2018lud}. It is easy to understand that
 most of the known neutron stars are pulsars. As of 1 December 2019,
 the ATNF Pulsar Catalogue~\cite{Manchester:2004bp,ATNF} (version 1.62)
 archived 2801 known pulsars. It is convenient to search the known pulsars
 in the ATNF Pulsar Catalogue~\cite{ATNF}. Since LIGO/Virgo S190814bv
 was well localized to a small area of 23 ${\rm deg}^2$ at $90\%$
 confidence~\cite{S190814bv,S190814bvGCN} centered on Right Ascension
 (J2000) RA = 00:50:37.5 (hms) and Declination (J2000)
 DEC = $-$25:16:57.371 (dms)~\cite{Dobie:2019ctw}, we search the known
 pulsars in the ATNF Pulsar Catalogue~\cite{ATNF} with this center (RA, DEC)
 and a radius of 3 degrees. There is only one known pulsar J0038$-$2501
 in this area (actually it is also the unique known pulsar even searching
 with a larger radius of 8 degrees) as of 1 December 2019. The unique known
 pulsar J0038$-$2501 in this direction was discovered by the Green Bank
 North Celestial Cap (GBNCC) Pulsar Survey recently, and its detailed
 information can be found in~\cite{Aloisi:2019ggk}. Its distance is
 given by~\cite{Aloisi:2019ggk}
 \be{eq1}
 d_{\rm ON}=320\,{\rm pc}\quad\quad
 {\rm or}\quad\quad d_{\rm ON}=600\,{\rm pc}\,,
 \ee
 according to NE2001~\cite{Cordes:2002wz} or YMW16~\cite{YMW16} models,
 respectively. Note that 1 pc = 3.261563777 light years. We will equally
 use these two distances in this work. Its direction in the sky
 reads~\cite{Aloisi:2019ggk}
 \be{eq2}
 \textrm{RA} = \textrm{00:38:10.264 (J2000, hms)}
 \textrm{\quad\quad and \quad\quad}
 \textrm{DEC} = -\textrm{25:01:30.73 (J2000, dms)} \,,
 \ee
 which is slightly outside the $90\%$ area of LIGO/Virgo S190814bv, but
 still inside the $2\sigma$ area. There are three possibilities: (1)
 Pulsar J0038$-$2501 might not be the one we want, and there might be
 another unknown neutron star in this direction as the mirror for GWs.
 (2) The true position of LIGO/Virgo S190814bv might be slightly outside
 the $90\%$ area given by~\cite{S190814bv,S190814bvGCN} due to measurement
 error. (3) The mirror image ``$\,\rm S^\prime\,$'' (i.e. LIGO/Virgo
 S190814bv) might slightly deviate from the direction of the GW mirror
 (i.e. the neutron star J0038$-$2501) due to the spherical shape of the
 neutron star (namely it is not a perfectly flat mirror). Taking the last
 two arguments into account, we adopt the pulsar J0038$-$2501 as the right
 mirror for GWs in this work.

Then, let us try to find the possible electromagnetic counterparts
 associated with the real NSBH merger ``\,S\,'' in the archive data
 before 14 September 2015. As mentioned above, short GRBs are the most
 promising electromagnetic counterparts associated with an NSBH merger.
 Fortunately, the first GRB was observed by the U.S. military satellites
 in the late 1960s (the discovery was declassified and published in the
 early 1970s)~\cite{GRBwiki}, and hence there exist rich archive data of
 GRBs which can be traced back to the early 1970s. Usually, the value of
 $T_{90}$ (the duration, in seconds, during which $90\%$ of the burst
 fluence was accumulated) separating short and long GRBs is
 $2\,{\rm s}$~\cite{Berger:2013jza}. A short GRB has a short duration
 $T_{90}< 2\,{\rm s}$.

So, we focus on the archive data of GRBs in this work. For a given GRB,
 its name is usually the date it was detected, and its trigger time is also
 recorded (for a few of GRBs their trigger times are absent, and then we can
 set them to be 00:00:00.000 UT for convenience). Thus, we can get the time
 delay $\Delta t$ between this GRB and LIGO/Virgo S190814bv triggered at
 21:10:39.012957 UT, 14 August 2019~\cite{S190814bv,S190814bvGCN}. Also,
 the position (RA, DEC) in the sky of this GRB is known, with an error
 radius. Thus, by using e.g. Astropy~\cite{Price-Whelan:2018hus,Astropy},
 we can also get the separate angle $\theta$ between this GRB and the GW
 mirror (namely the neutron star J0038$-$2501) whose position in the
 sky is given by Eq.~(\ref{eq2}). As mentioned above (see
 also~\cite{Wei:2019sqw}), from the left panel of Fig.~\ref{fig1}, if
 this GRB is associated with the real NSBH merger ``\,S\,'', we have
 a simple relation
 \be{eq3}
 c\Delta t\simeq\left(1-\cos\theta\right)d_{\rm ON}\,,
 \ee
 due to the huge distance $d_{\rm OS}\gg d_{\rm ON}$. The distance of
 the GW mirror (i.e. the neutron star J0038$-$2501) is already given by
 Eq.~(\ref{eq1}). Obviously, it is convenient to use $d_{\rm ON}$ and
 $\Delta t$ in units of light years and years, respectively. Note that
 in Eq.~(\ref{eq3}), we do not need to know the distance or the redshift
 of the given GRB. This is a great advantage in fact. We can
 check the relation in Eq.~(\ref{eq3}) to see whether a given GRB is
 indeed associated with the real NSBH merger ``\,S\,''. However, the
 position (RA, DEC) in the sky of the given GRB is not exactly
 measured, and its $1\sigma$ uncertainty in the position is usually
 characterized by an error radius (which is also given in the GRB data
 usually). So, we introduce a new quantity to describe the deviation
 between the central position of this GRB and the real
 NSBH merger ``\,S\,'', namely
 \be{eq4}
 \Delta\theta\equiv\left|\,\arccos\left(1-c\Delta t/d_{\rm ON}\right)
 -\theta\,\right|\,,
 \ee
 where $|x|$ denotes the absolute value of any $x$. Obviously,
 $\Delta\theta\simeq 0$ for the GRB associated with the real
 NSBH merger ``\,S\,''. Due to the uncertainty in the position of GRB,
 if $\Delta\theta$ is less than the corresponding error radius, this
 GRB is a possible electromagnetic counterpart associated with the real
 NSBH merger ``\,S\,''. In other words, this GRB can satisfy the relation
 in Eq.~(\ref{eq3}) within its $1\sigma$ uncertainty area in the position.
 Of course, we should calculate two $\Delta\theta$ for the two distances
 $d_{\rm ON}$ of the GW mirror (namely the neutron star J0038$-$2501) in
 Eq.~(\ref{eq1}) derived from the NE2001 and YMW16 models, respectively.

Clearly, it is not necessary to scan every GRB detected before
 14 September 2015 in the whole sky. Noting Eq.~(\ref{eq3}),
 for a separate angle $\theta>25^\circ$ and $d_{\rm ON}>1000$ light
 years, the time delay $\Delta t > 93.7$ years, far beyond the time
 when the first GRB was detected in the late 1960s. Therefore, it is
 enough to scan the GRBs detected before 14 September 2015 in
 a smaller area centered around the neutron star J0038$-$2501, namely
 \be{eq5}
 {\rm 22\,h}<{\rm RA}<{\rm 03\,h}\quad\quad {\rm and}\quad\quad
 -50\,{\rm deg} <{\rm DEC}< 0\,{\rm deg}\,.
 \ee

Most GRBs were detected by the telescopes Fermi, Swift, INTEGRAL, BATSE,
 AGILE, HETE, Konus-Wind, Beppo-SAX, MAXI, IPN and so on. Let us consider
 the Fermi GBM GRBs~\cite{Fermi,Bhat:2016odd} at first. The well-known
 Fermi Gamma-ray Space Telescope was launched on 11 June 2008. In fact, the
 first electromagnetic signal of the well-known BNS merger GW170817 was
 detected by Fermi GBM~\cite{GBM:2017lvd,Monitor:2017mdv,GRB170817A},
 and it sent the first GCN circular of GW170817 before
 LIGO/Virgo~\cite{G298048}. As of 1 December 2019, Fermi GBM detected
 2695 GRBs~\cite{Fermiquery}. It is easy to find all Fermi GBM GRBs in
 the area defined by Eq.~(\ref{eq5}) before 14 September 2015 by using
 the online query form~\cite{Fermiquery} (see also~\cite{heasarc}). There
 are 143 short and long GRBs on hand. Then, we calculate the corresponding
 $\Delta t$, $\theta$ and $\Delta\theta$ for these 143 GRBs. Finally, we
 find 9 short GRBs (namely $T_{90}<2\,{\rm s}$ by taking its error into
 account) which can satisfy the relation in Eq.~(\ref{eq3}) within its
 $1\sigma$ uncertainty area in the position (namely $\Delta\theta<$ error
 radius), and we present them in Table~\ref{tab1}. In these 9 short
 GRBs, GRB120314412 can only be a possible electromagnetic counterpart
 for the NE2001 case but not for the YMW16 case, while the other 8 short
 GRBs can be the possible electromagnetic counterparts for both the
 NE2001 and YMW16 cases. Note that a (too) large error radius means that
 the localization of the GRB is not well determined. So, we classify the
 last 5 short GRBs as ``\,Silver\,'' candidates because of their large
 error radii and/or large $\Delta\theta$, and classify the first 4 short
 GRBs as ``\,Gold\,'' candidates. It is worth noting that such
 a classification is subtle/subjective and just for reference.
 In principle, all these 9 short GRBs can be the possible
 electromagnetic counterparts associated with the real NSBH
 merger ``\,S\,'' corresponding to LIGO/Virgo S190814bv.

Next, we turn to the well-known Swift Gamma-Ray Burst Mission (renamed
 the Neil Gehrels Swift Observatory on 10 January 2018)~\cite{Swift,
 Swiftcat,Swiftgrb}, which was launched on 20 November 2004. Actually,
 in the extensive follow-up observing campaign of the well-known BNS
 merger GW170817~\cite{G298048}, Swift caught the first ultraviolet
 light from GW170817~\cite{Evans:2017mmy}. It has found 1332 GRBs as
 of 1 December 2019~\cite{Swift}. The error radii of Swift GRBs are
 commonly a few arcmins. There are 73 Swift short and long GRBs in the
 area defined by Eq.~(\ref{eq5}) before 14 September 2015~\cite{Swift}.
 Unfortunately, all these 73 GRBs are not electromagnetic counterparts
 associated with the real NSBH merger ``\,S\,'' corresponding
 to LIGO/Virgo S190814bv, because their $\Delta\theta$ are all
 larger than the corresponding error radii in the position.


 \begin{sidewaystable}[tbp] 
 \renewcommand{\arraystretch}{1.7}
 \begin{center}
 \begin{tabular}{ccccccccccc} \hline\hline
 GRB name  & GRB name  & RA &  DEC  & Error radius & \ $T_{90}$ \ & \ \ \quad $\Delta t$ \quad\ \quad
 & $\theta$  & $\Delta\theta$ (NE2001) \quad & $\Delta\theta$ (YMW16) \quad & Candidate \\[-2mm]
 (GRByymmddfff) & \ (GRByymmddx)\ & \ (J2000, hms)\ & \ (J2000, dms)\ & (deg) & (s) & (year) & (deg) & (deg) & (deg) & (G/S) \\ \hline
 GRB100107074 & GRB100107A & 00:25:14.4   & $-$21:14:24   & 5.97    & $0.576\pm 0.465$  &  9.601    & 4.813	    & 2.965	    & 0.865   & Gold \\
 GRB141122087 &            & 00:38:50.4   & $-$20:01:12   & 10.90   & $1.280\pm 0.945$  & 4.728     & 5.008	    & 0.448	    & 1.024   & Gold \\
 GRB101214748 & GRB101214A & 00:02:45.6   & $-$28:16:12	  & 5.56    & $2.240\pm 2.084$  & 8.666     & 8.549	    & 1.160	    & 3.155   & Gold \\
 GRB090108322 & GRB090108B & 00:01:36.0   & $-$32:54:00	  & 8.30    & $0.192\pm 0.143$  & 10.597    & 11.215	& 3.043	    & 5.250   & Gold \\ \hline
 GRB081213173 & GRB081213  & 00:51:36.0   & $-$33:54:00	  & 13.20   & $0.256\pm 0.286$  & 10.669    & 9.342	    & 1.143	    & 3.356   & Silver \\
 GRB120524134 & GRB120524A & 23:52:36.0   & $-$15:36:36	  & 10.45   & $0.704\pm 0.466$  & 7.224     & 14.226	& 7.481	    & 9.301   & Silver \\
 GRB120314412 & GRB120314A & 01:11:33.6   & $-$48:43:48	  & 17.82   & $1.280\pm 1.086$  & 7.418     & 24.592	& 17.756	& 19.601  & Silver \\
 GRB120926753 &            & 01:38:26.4   & $-$45:34:48	  & 21.32   & $3.072\pm 2.064$  & 6.881     & 23.860	& 17.278	& 19.054  & Silver \\
 GRB140109877 &            & 01:36:21.6   & $-$25:03:00	  & 37.45   & $3.328\pm 2.560$  & 5.593     & 13.174	& 7.239	    & 8.841   & Silver \\
 \hline\hline
 \end{tabular}
 \end{center}
 \caption{\label{tab1} The Fermi GBM short GRBs probably associated with
 the real NSBH merger ``\,S\,'' corresponding to LIGO/Virgo S190814bv.
 We classify them as ``\,Gold\,'' candidates or ``\,Silver\,''
 candidates. Note that some of these GRBs have not been reported in GCN,
 but can be found in the published Fermi GBM GRB Catalog~\cite{Fermi,
 Bhat:2016odd}. These results are for case (I). See the text for details.}
 \end{sidewaystable}


Obviously, it is not convenient to check every telescope for GRBs one
 by one (with the exception of telescopes like Fermi and Swift which
 have their own remarkable GRB databases). Fortunately and gratefully,
 there are several online GRB databases created and maintained
 by volunteers, such as GRBOX~\cite{GRBOX}, Jochen Greiner's GRB
 table~\cite{grbgen}, GRBlog~\cite{GRBlog} and so on. Some of
 them have been discontinued, but some are still living. They usually
 collect the GRB information from GCN circulars, and hence they
 contain numerous GRBs from various telescopes, but the GRB parameters
 are preliminary and final results should be found in the published
 GRB Catalog of the corresponding telescope. On the other hand, they
 might also miss some GRBs which have not been reported in GCN. In this
 work, we choose to use GRBOX~\cite{GRBOX}, since it has a user-friendly
 interface and we only need GRBs before 14 September 2015. In
 GRBOX~\cite{GRBOX}, it is easy to get 149 short and long GRBs in the
 area defined by Eq.~(\ref{eq5}) before 14 September 2015, while we have
 excluded 29 GRBs without RA and DEC from the original 178 GRBs. Note that
 the pre-1990 GRBs (including 4 GRBs in the 1970s) have coordinates
 reported in B1950, and we have transformed them into J2000 by using the
 NED Coordinate Transformation~\cite{NEDtrans}. Again, we calculate
 the corresponding $\Delta t$, $\theta$ and $\Delta\theta$ for these
 149 GRBs. Since $T_{90}$ in GRBOX is given without error information,
 we relax it to $T_{90}<5\,{\rm s}$ for the possible short GRBs, and
 also take GRBs whose $T_{90}$ are not available into account. So, we
 find 4 GRBs with $T_{90}<5\,{\rm s}$ (or n/a) and error radius
 $>\Delta\theta$. They are presented in Table~\ref{tab2}. As mentioned
 above, they were collected from GCN circulars, and hence their GRB
 parameters are preliminary. We should check them in the published GRB
 Catalog of the corresponding telescope. In fact, GRB100415A was found by
 MAXI~\cite{Serino:2014wza}. We check GRB100415A in~\cite{Serino:2014wza}
 and find that its $T_{90}\geq T_d = 23.4\,{\rm s}$. So, it should be
 excluded because it is a long GRB in fact. Similarly,
 GRB090426B is actually the Ferimi GBM GRB090426066, whose
 final $T_{90}=16.128\pm 5.152\,{\rm s}$~\cite{Bhat:2016odd}, and should
 also be excluded since it is a long GRB instead. On the other hand,
 GRB090108B and GRB081213 are also Fermi GBM GRBs~\cite{Bhat:2016odd},
 which are indeed short GRBs probably associated with the real NSBH
 merger ``\,S\,'' corresponding to LIGO/Virgo S190814bv, as shown in
 Table~\ref{tab1} (and the results of these two GRBs in Table~\ref{tab1}
 should prevail).


 \begin{table}[tb] 
 \renewcommand{\arraystretch}{1.7}
 \begin{center}
 \begin{tabular}{ccccccccccc} \hline\hline
 GRB name & \ \ $T_{90}$ \ & RA &  DEC  & \ Error radius &  $\Delta t$
 & $\theta$  & $\Delta\theta$ (NE2001) \quad & $\Delta\theta$ (YMW16) \\[-2mm]
 (yymmddx) & \ \ (s) \  & \ (J2000, deg)\ & \ (J2000, deg)\ & (deg) & \ (year) \ & \ \ (deg) \ \ & (deg) & (deg) \\ \hline
 090108B	& 0.8	& 3.75	& $-32.2$	& 6.4 	      & 10.597	& 8.790	  & 0.619	& 2.825 \\
 081213		& n/a   & 25.5	& $-35.3$	& 12.5 	      & 10.669	& 17.163  & 8.964	& 11.178 \\
 100415A	& n/a   & 7.7	& $-16.46$	& $\sim 2$ 	  & 9.333	& 8.736	  & 1.069	& 3.139 \\
 090426B	& 3.8	& 17.5	& $-19.2$	& 18.1 	      & 10.302	& 9.391	  & 1.334	& 3.510 \\
 \hline\hline
 \end{tabular}
 \end{center}
 \caption{\label{tab2} The GRBOX GRBs in the area defined by
 Eq.~(\ref{eq5}) before 14 September 2015, with $T_{90}<5\,{\rm s}$
 (or n/a) and error radius $>\Delta\theta$. These results
 are for case (I). See the text for details.}
 \end{table}



 \begin{sidewaystable}[tbp] 
 \renewcommand{\arraystretch}{1.7}
 \begin{center}
 \begin{tabular}{ccccccccccc} \hline\hline
 GRB name  & GRB name  & RA &  DEC  & Error radius & \ $T_{90}$ \ & \ \ \quad $\Delta t$ \quad\ \quad
 & $\theta$  & Candidate \\[-2mm]
 (GRByymmddfff) & \ (GRByymmddx)\ & \ (J2000, hms)\ & \ (J2000, dms)\ & (deg) & (s) & (year) & \ (deg) \ \quad &  (G/S)  \\ \hline
 GRB100107074 & GRB100107A & 00:25:14.4   & $-$21:14:24   & 5.97    & $0.576\pm 0.465$  &  9.601    & 7.093	    & Gold \\
 GRB141122087 &            & 00:38:50.4   & $-$20:01:12   & 10.90   & $1.280\pm 0.945$  & 4.728     & 5.923	    & Gold \\ \hline
 GRB081213173 & GRB081213  & 00:51:36.0   & $-$33:54:00	  & 13.20   & $0.256\pm 0.286$  & 10.669    & 8.620	    & Silver \\
 GRB140109877 &            & 01:36:21.6   & $-$25:03:00	  & 37.45   & $3.328\pm 2.560$  & 5.593     & 10.348	& Silver \\
 GRB120926753 &            & 01:38:26.4   & $-$45:34:48	  & 21.32   & $3.072\pm 2.064$  & 6.881     & 22.455	& Silver \\
 \hline\hline
 \end{tabular}
 \end{center}
 \caption{\label{tab3} The Fermi GBM short GRBs probably associated with
 the real NSBH merger ``\,S\,'' corresponding to LIGO/Virgo S190814bv.
 We classify them as ``\,Gold\,'' candidates or ``\,Silver\,''
 candidates. Note that some of these GRBs have not been reported in GCN,
 but can be found in the published Fermi GBM GRB Catalog~\cite{Fermi,
 Bhat:2016odd}. These results are for cases (II) and (III),
 and hence the separate angles $\theta$ are different from the ones in
 Table~\ref{tab1} for case (I). See the text for details.}
 \end{sidewaystable}


 \vspace{-3mm}  


\subsection{Cases (II) and (III)}\label{sec3b}

Here, we turn to cases (II) and (III), in which the neutron
 star ``\,N\,'' as the mirror for GWs is located in the same host galaxy
 of the real NSBH merger ``\,S\,'' corresponding to LIGO/Virgo
 S190814bv, as shown in the middle and right panels of Fig.~\ref{fig1}.
 In these cases, the neutron star ``\,N\,'' is certainly unknown. Only
 the GW mirror image ``\,$\rm S^\prime$\,'' is known, namely LIGO/Virgo
 S190814bv at a distance of $267\pm 52\,{\rm Mpc}$~\cite{S190814bv,
 S190814bvNotice,S190814bvGCN}. It was well localized to a small area
 of 23 ${\rm deg}^2$ at $90\%$ confidence~\cite{S190814bv,S190814bvGCN}
 centered on~\cite{Dobie:2019ctw}
 \be{eq6}
 \textrm{RA} = \textrm{00:50:37.5 (J2000, hms)}
 \textrm{\quad\quad and \quad\quad}
 \textrm{DEC} = -\textrm{25:16:57.371 (J2000, dms)} \,.
 \ee
 As mentioned above, the time delay
 $c\Delta t\simeq\left(1-\cos\phi\right)d_{\rm SN}$ and
 $c\Delta t\simeq\left(1+\cos\psi\right)d_{\rm SN}$ in cases (II) and
 (III), respectively. The separate angle $\theta$ between the real NSBH
 merger ``\,S\,'' and the GW mirror image ``\,$\rm S^\prime$\,'' (namely
 LIGO/Virgo S190814bv) must be very close to 0, due to the huge distance
 $d_{\rm OS}\simeq d_{\rm OS^\prime}=267\pm 52\,{\rm Mpc}$. In fact,
 $\theta$ is on the order of
 $d_{\rm SN}/d_{\rm OS}\sim c\Delta t/d_{\rm OS}\sim 10^{-8}$ or $10^{-7}$.
 So, if a GRB is associated with the real NSBH merger ``\,S\,'', the
 separate angle $\theta$ between this GRB and LIGO/Virgo S190814bv should
 be $\theta\to 0$. However, as mentioned above, the position (RA, DEC)
 in the sky of a given GRB is not exactly measured, and its $1\sigma$
 uncertainty in the position is usually characterized by an error radius.
 On the other hand, the position of LIGO/Virgo S190814bv has also not been
 exactly measured in fact. Its true position in the sky is somewhere in an
 area of 23 ${\rm deg}^2$ at $90\%$ confidence~\cite{S190814bv,S190814bvGCN}
 centered on Eq.~(\ref{eq6}), as mentioned above. So, if the $1\sigma$
 uncertainty area in the position of a given GRB overlaps with the one of
 LIGO/Virgo S190814bv, this GRB might be associated with the real NSBH
 merger ``\,S\,'' corresponding to LIGO/Virgo S190814bv. In this case,
 the separate angle $\theta$ between the central positions of this GRB
 and LIGO/Virgo S190814bv can be not close to 0. For convenience, the
 $90\%$ uncertainty area in position of LIGO/Virgo S190814bv can be
 conservatively approximated to an area centered on Eq.~(\ref{eq6})
 with an error radius $\sim 2.5\,{\rm deg}$. Therefore, if the separate
 angle $\theta$ is less than the corresponding error radius of a given GRB
 plus $2.5\,{\rm deg}$, this GRB might be associated with the real NSBH
 merger ``\,S\,'' corresponding to LIGO/Virgo S190814bv.

Similar to Sec.~\ref{sec3a}, it is not necessary to scan all GRBs in
 the whole sky. Instead, it is also enough to scan the GRBs detected
 before 14 September 2015 in a smaller area defined by Eq.~(\ref{eq5}).
 We consider the Fermi GBM GRBs~\cite{Fermi,Bhat:2016odd} at first, and
 present the results in Table~\ref{tab3}. There are 5 short GRBs associated
 with the real NSBH merger ``\,S\,'' corresponding to LIGO/Virgo S190814bv
 for cases (II) and (III). Note that a (too) large error radius means
 that the localization of the GRB is not well determined. So, we classify
 the last 3 short GRBs as ``\,Silver\,'' candidates because of their
 large error radii and/or large $\theta$, and classify the first 2 short
 GRBs as ``\,Gold\,'' candidates. It is worth noting that these 5 short
 GRBs are also included in Table~\ref{tab1}. That is, they are
 the possible electromagnetic counterparts associated with the real
 NSBH merger ``\,S\,'' corresponding to LIGO/Virgo S190814bv in
 all cases (I), (II) and (III).

Next, we turn to other GRB databases. Let us consider the Swift
 GRBs~\cite{Swift}. Unfortunately, none of them could be the
 electromagnetic counterparts associated with the real NSBH
 merger ``\,S\,'' corresponding to LIGO/Virgo S190814bv. Similar to
 Sec.~\ref{sec3a}, we also consider the GRBOX GRBs~\cite{GRBOX}, and
 find only two GRBs (GRB090426B and GRB081213) might be the
 electromagnetic counterparts for cases (II) and (III). However, as
 mentioned at the end of Sec.~\ref{sec3a}, GRB090426B is a long GRB
 in fact, and it should be excluded. So, only GRB081213 is the
 possible electromagnetic counterpart associated with the real NSBH
 merger ``\,S\,'' corresponding to LIGO/Virgo S190814bv for cases (II)
 and (III). Noting that GRB081213 has also been presented in
 Table~\ref{tab3}, we do not show it again.


\section{Summary}\label{sec4}

LIGO/Virgo S190814bv is the first high-probability NSBH merger
 candidate, whose GWs triggered LIGO/Virgo detectors at 21:10:39.012957
 UT, 14 August 2019~\cite{S190814bv,S190814bvNotice,S190814bvGCN}. It
 has a probability $>99\%$ of \mbox{being} an NSBH merger, with a low
 false alarm rate (FAR) of 1 per 1.559e+25 years. For an NSBH merger,
 electromagnetic counterparts (especially short GRBs) are generally
 expected. However, no electromagnetic counterpart has been found in the
 extensive follow-up observing campaign~\cite{S190814bvGCN}. In the present
 work, we propose a novel explanation to this null result. In our scenario,
 LIGO/Virgo S190814bv is just a GW mirror image of the real NSBH merger
 which should be detected many years before 14 September 2015, but at
 that time we had no ability to detect its GW signals. The electromagnetic
 counterparts associated with the real NSBH merger ``\,S\,'' corresponding
 to LIGO/Virgo S190814bv should be found in the archive data before 14
 September 2015. In this work, we indeed find 9 short GRBs as the possible
 electromagnetic counterparts. The names of all 9 short GRBs can be found
 in Table~\ref{tab1} (note that the names of 5 short GRBs in
 Table~\ref{tab3} are also included in Table~\ref{tab1}).


\section*{ACKNOWLEDGEMENTS}
We thank the anonymous referee for quite useful comments and
 suggestions, which helped us to improve this work. We are
 grateful to Profs. Rong-Gen~Cai, Shuang~Nan~Zhang, Bobing~Wu,
 Jian-Min~Wang, Zong-Kuan~Guo, Zhoujian~Cao, Wen~Zhao, Bin~Hu,
 Hongsheng~Zhang, Yi~Zhang and Lijun~Gou for helpful discussions. We
 also thank Da-Chun~Qiang, Zhong-Xi~Yu, Hua-Kai~Deng, and Shu-Ling~Li
 for kind help and discussions. This work was supported in part by NSFC
 under Grants No.~11975046 and No.~11575022.

\newpage 

\renewcommand{\baselinestretch}{1.05}


\end{document}